\documentclass[useAMS,usenatbib]{mnras}
\usepackage{graphicx}
\usepackage{float}
\usepackage{newtxtext,newtxmath}
\usepackage[outdir=./]{epstopdf}
\usepackage{tablefootnote}
\DeclareGraphicsExtensions{.eps}
\DeclareGraphicsExtensions{.pdf}
\usepackage[dvipsnames]{xcolor}
\usepackage{soul}

\title[Frequency of gaseous discs at white dwarfs]{The frequency of gaseous debris discs around white dwarfs}

\author[Manser et al.]{Christopher J. Manser\,$^1$\,{\Huge \footnotemark},
Boris T. G\"ansicke\,$^{1,2}$,
Nicola Pietro Gentile Fusillo\,$^{1,3}$,
\newauthor
Richard Ashley\,$^1$,
Elm\'e Breedt\,$^4$,
Mark Hollands\,$^1$,
Paula Izquierdo\,$^{5,6}$,
Ingrid Pelisoli\,$^{7,8}$\\
$^{1}$ Department of Physics, University of Warwick, Coventry CV4 7AL,
UK \\
$^{2}$ Centre for Exoplanets and Habitability, University of Warwick, Coventry, UK \\
$^{3}$ European Southern Observatory, Karl-Schwarzschild-Stra{\ss}e 2, 85748 Garching bei M\"{u}nchen, Germany \\
$^{4}$ Institute of Astronomy, University of Cambridge, Cambridge, CB3 0HA, UK \\
$^{5}$ Instituto de Astrof\'isica de Canarias, E-38205 La Laguna, Tenerife, Spain \\
$^{6}$ Departamento de Astrof\'isica, Universidad de La Laguna, E-38206 La Laguna, Tenerife, Spain \\
$^{7}$ Institut f\"ur Physik und Astronomie, Universit\"atsstandort Golm, Karl-Liebknecht-Str. 24/25, 14467 Potsdam, Germany \\
$^{8}$ Instituto de F\'{i}sica, Universidade Federal do Rio Grande do Sul, 91501-900, Porto-Alegre, RS, Brazil
}

\begin{document}

\date{Accepted 20XX. Received 20XX; in original form 20XX}

\pagerange{\pageref{firstpage}--\pageref{lastpage}} \pubyear{20XX}

\maketitle

\label{firstpage}

\begin{abstract}
1\,--\,3\,per\,cent of white dwarfs are orbited by planetary dusty debris detectable as infrared emission in excess above the white dwarf flux. In a rare subset of these systems, a gaseous disc component is also detected via emission lines of the Ca\,{\textsc{ii}} 8600\,\AA\ triplet, broadened by the Keplerian velocity of the disc. We present the first statistical study of the fraction of debris discs containing detectable amounts of gas in emission at white dwarfs within a magnitude and signal-to-noise limited sample. We select 7705 single white dwarfs spectroscopically observed by the Sloan Digital Sky Survey (SDSS) and \textit{Gaia} with magnitudes $g$\,$\leq$\,19. We identify five gaseous disc hosts, all of which have been previously discovered. We calculate the occurrence rate of a white dwarf hosting a debris disc detectable via Ca\,{\textsc{ii}} emission lines as 0.067\,$\pm$\,$_{0.025}^{0.042}$\,per\,cent. This corresponds to an occurrence rate for a dusty debris disc to have an observable gaseous component in emission as 4\,$\pm$\,$_{2}^{4}$\,per\,cent. Given that variability is a common feature of the emission profiles of gaseous debris discs, and the recent detection of a planetesimal orbiting within the disc of SDSS\,J122859.93+104032.9, we propose that gaseous components are tracers for the presence of planetesimals embedded in the discs and outline a qualitative model. We also present spectroscopy of the Ca\,{\textsc{ii}} triplet 8600\,\AA\ region for 20 white dwarfs hosting dusty debris discs in an attempt to identify gaseous emission. We do not detect any gaseous components in these 20 systems, consistent with the occurrence rate that we calculated.
\end{abstract}


\footnotetext{E-mail: c.j.manser92@googlemail.com}

\section{Introduction}
It is well established that remnant planetary systems exist around white dwarfs; betrayed by the photospheric metal contamination of a large number of these stellar remnants \citep{zuckermanetal10-1, koesteretal14-1, hollandsetal17-1, barstowetal14-1, schreiberetal19-1}. The time-scale on which material sinks through the atmosphere of a warm, hydrogen-rich (DA) white dwarf and becomes unobservable is on the order of days to years, and therefore this atmospheric pollution needs to be sustained by ongoing accretion to remain detectable \citep{koester09-1, wyattetal14-1}. The generally accepted framework explaining the pollution of white dwarfs below $\simeq$\,30\,000\,K is the tidal disruption of a planetesimal \citep{debes+sigurdsson02-1, jura03-1}, which forms a dusty debris disc that is subsequently accreted by the white dwarf \citep{verasetal14-1, verasetal15-1, malamud+perets19-1, malamud+perets19-2}. The thermal infrared emission of such dusty debris discs \citep{grahametal90-1} has been detected at 37 metal polluted white dwarfs as an infrared excess above the white dwarf continuum \citep{zuckerman+becklin87-1, vanderburgetal15-1, rocchettoetal15-1, barberetal16-1, dennihyetal16-1}. In a subset of these systems, gaseous debris has also been observed that is co-orbital with the dust \citep{melisetal10-1}. This gaseous material has so far been detected in seven systems \citep{gaensickeetal06-3, gaensickeetal07-1, gaensickeetal08-1, gaensicke11-1, farihietal12-1, melisetal12-1, wilsonetal14-1}, via the Doppler-broadened, double-peaked line emission of the Ca\,{\textsc{ii}} 8600\,\AA\ triplet, which results from the Keplerian rotation of a flat disc \citep{horne+marsh86-1} that is photoionised by the white dwarf \citep{melisetal10-1, kinnear11, gaensickeetal19-1}.

In addition, five white dwarfs have been found to host gas that is detected via circumstellar absorption features \citep{koester+wilken06-1, debesetal12-2, gaensickeetal12-1, vennes+kawka13-1, koesteretal14-1, vanderburgetal15-1, xuetal16-1}. However, the spatial extent of the circumstellar gas has so far only been constrained for one of these five systems (WD\,1145+017, \citealt{hallakounetal17-1, cauleyetal18-1, izquierdoetal18-1, xuetal19-1, fortinetal19-1}), which does not appear to be co-orbital with the dusty debris. The detection of circumstellar absorption features is not correlated with the presence of emission lines from gaseous disc components, suggesting the mechanisms that generates both phenomena are independent. In this work, we therefore focus on the systems showing double-peaked emission lines from a circumstellar gaseous disc, and henceforth refer to those simply as \textit{gaseous discs}.

The gas producing emission co-orbits with the dust and several mechanisms to generate gaseous components have been proposed, including runaway gas production from dust sublimating at the inner edge of the debris disc \citep{rafikov11-2, metzgeretal12-1}, collisional cascades of solid material being ground into dust and gas \citep{kenyon+bromley17-1, kenyon+bromley17-2}, and planetesimals stirring up the disc \citep{manseretal19-1}. A knowledge of the occurrence rate of the gaseous components in emission to these discs will allow constraints to be placed on the formation and evolution of debris discs at white dwarfs \citep{rafikov11-2, metzgeretal12-1}, and possibly the incidence of closely orbiting planetesimals at white dwarfs \citep{manseretal19-1}.

Throughout the last decade sufficiently large and unbiased samples of white dwarfs have been established to determine the occurrence rate of (i) observable dust discs around, and (ii) photospheric pollution of white dwarfs, as 1\,--\,3\,per\,cent and 25\,--\,50\,per\,cent, respectively \citep{zuckermanetal03-1, farihietal09-1, zuckermanetal10-1, girvenetal11-1, koesteretal14-1, rocchettoetal15-1, wilsonetal19-1, rebassa-mansergasetal19-1}. However, the prevalence of detectable gaseous components in emission to these debris discs is so far unconstrained. Simply taking the number of the currently known gaseous components (seven) and dusty debris discs (37) suggests an occurrence rate of $\simeq$\,19\,per\,cent, which is, however erroneous due to the different observational biases in the detection of the dusty and gaseous components.

In this paper we present the first statistical study of the frequency of debris discs that host gaseous emission, using a magnitude and signal-to-noise (S/N) limited \textit{Gaia}/Sloan Digital Sky Survey (SDSS) sample of 7705 white dwarfs. We searched the SDSS spectroscopy for Ca\,{\textsc{ii}} triplet emission and determined the percentage of white dwarfs that host an observable gaseous debris disc as 0.067\,$\pm$\,$_{0.025}^{0.042}$\,per\,cent. This corresponds to an occurrence rate for a dusty debris disc detected as an infrared excess to also have an observable gaseous component as 4\,$\pm$\,$_{2}^{4}$\,per\,cent. We also present spectroscopy of the Ca\,{\textsc{ii}} triplet region for 20 white dwarfs known to host dusty discs, some of which have no published observations of the Ca\,{\textsc{ii}} triplet, and we do not detect a gaseous disc component in any of these systems. 
 
\section{Searching for gaseous debris discs in SDSS spectroscopy}
\label{sec:gasdiscsearch}

\begin{figure}
\centerline{\includegraphics[width=1\columnwidth]{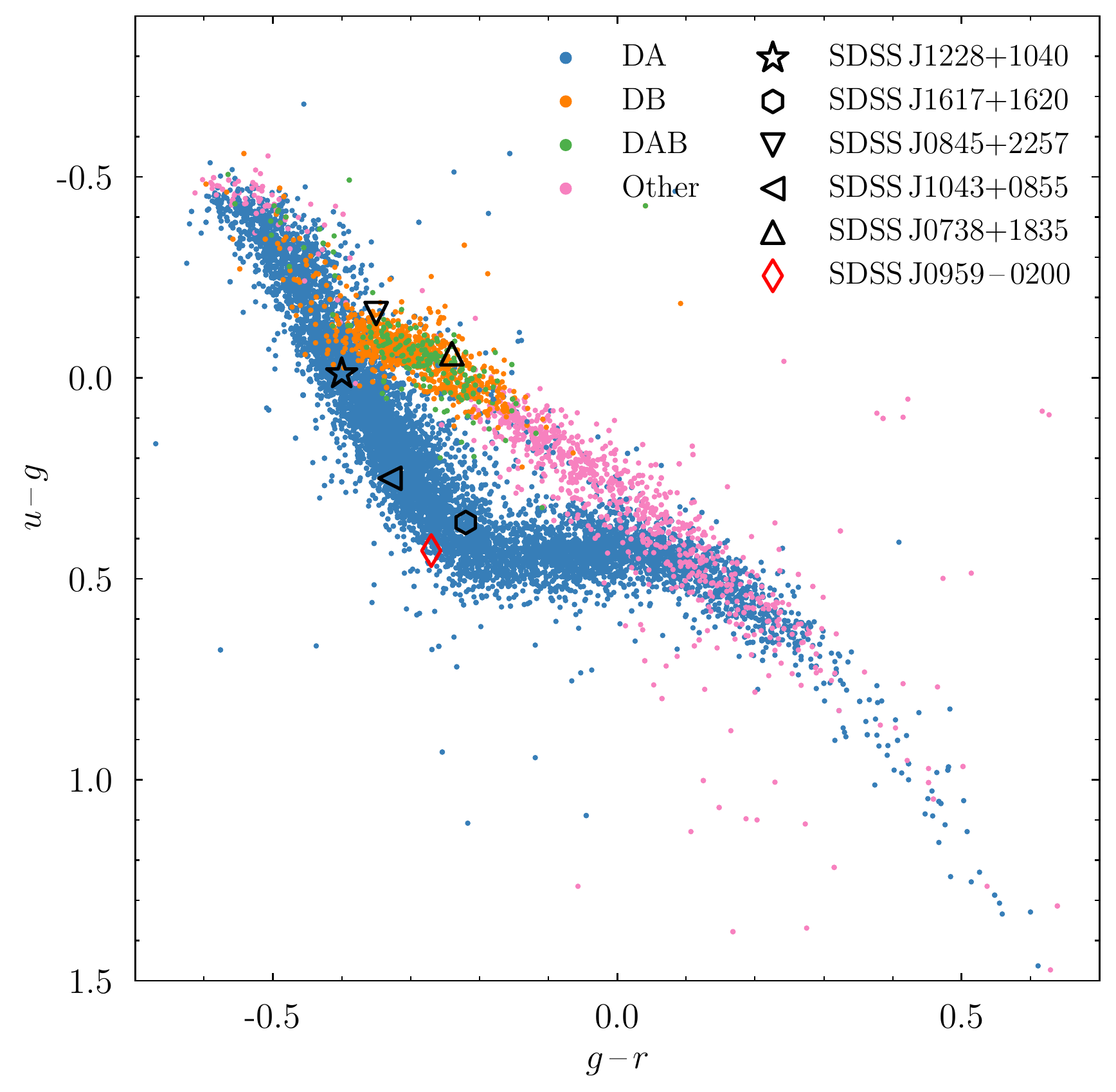}}
\caption{\label{f-colourplot} $u$--$g$, $g$--$r$ colour-colour diagram of our magnitude-limited ($g\leq19$) white dwarf sample\protect\footnotemark[1]\ obtained from \protect\cite{fusilloetal19-1}. The main spectral types of white dwarfs are labelled by colour, and the five white dwarfs with a gaseous component to their debris disc contained in this sample are shown in black. SDSS\,J0959--0200 is included for reference in red. See Table\,\ref{t-logwht} and text for additional information.}
\end{figure}

The SDSS has been taking multi-band photometry and multi-fibre spectroscopy since 2000, using a 2.5\,m telescope located at the Apache Point Observatory in New Mexico \citep{gunnetal06-1}. \cite{fusilloetal19-1} published a sample of spectroscopically observed white dwarfs combining the astrometry of \textit{Gaia} and the large number of spectra obtained by SDSS \citep{gaiaetal18-1}. We extract from this \textit{Gaia}/SDSS catalogue a magnitude limited sample of single white dwarfs with $g$\,$\leq$\,19 that contains 9374 white dwarfs with 14\,040 spectra ($\simeq$\,38\,per\,cent of white dwarfs in the sample have more than one spectrum). Figure\,\ref{f-colourplot} shows this sample of white dwarfs in the $u$--$g$ vs. $g$-$r$ colour-colour space.

\footnotetext[1]{A small number of white dwarfs (48) are not visible on this plot, which mainly comprise of the cool, metal polluted white dwarfs that are found below the main-sequence in the $u$--$g$ vs. $g$-$r$ colour-colour space, i.e. with $u$--$g$\,$\geq$\,1.5 \citep{koesteretal11-1, hollandsetal17-1}. This is due to the large amount of flux blocked in the $u$ band due to absorption lines from calcium, iron, magnesium, etc. (see fig.\,1 of \citealt{hollandsetal18-1}).}

\begin{figure*}
\centerline{\includegraphics[width=2\columnwidth]{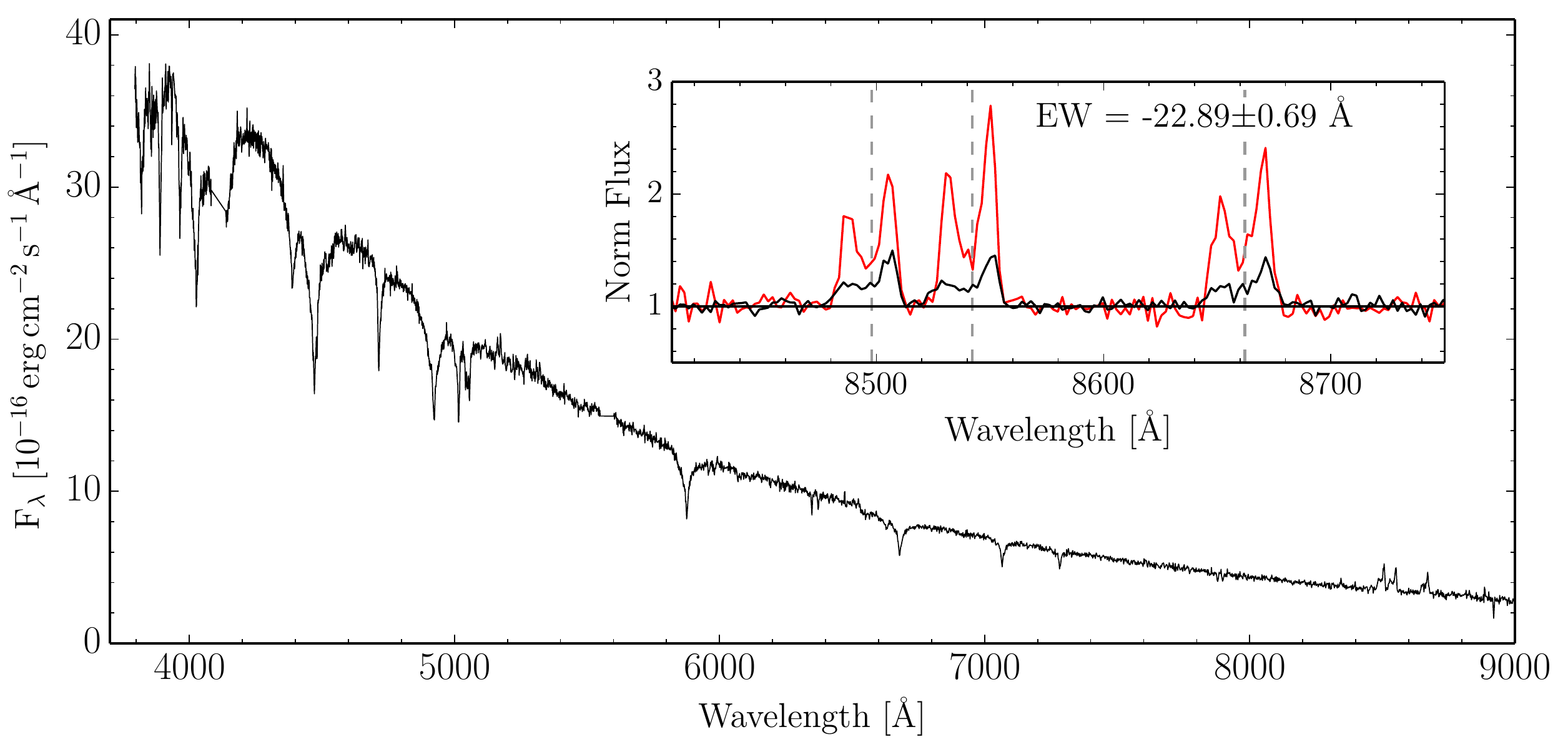}}
\caption{\label{f-normplot} The SDSS spectrum of SDSS\,J0845+2257 (black) is shown as an example to illustrate our search for gas discs. The inset shows the Ca\,{\textsc{ii}} triplet region with the rest wavelengths (dashed) marked and the combined equivalent width of the triplet given. The prototypical emission profiles of SDSS\,J1228+1040 (red) are also shown for comparison. All SDSS spectra of the gaseous debris discs in our \textit{Gaia}/SDSS sample are shown in Figure\,\ref{f-normplot_appendix}.}
\end{figure*} 

\begin{figure*}
\centerline{\includegraphics[width=1\columnwidth]{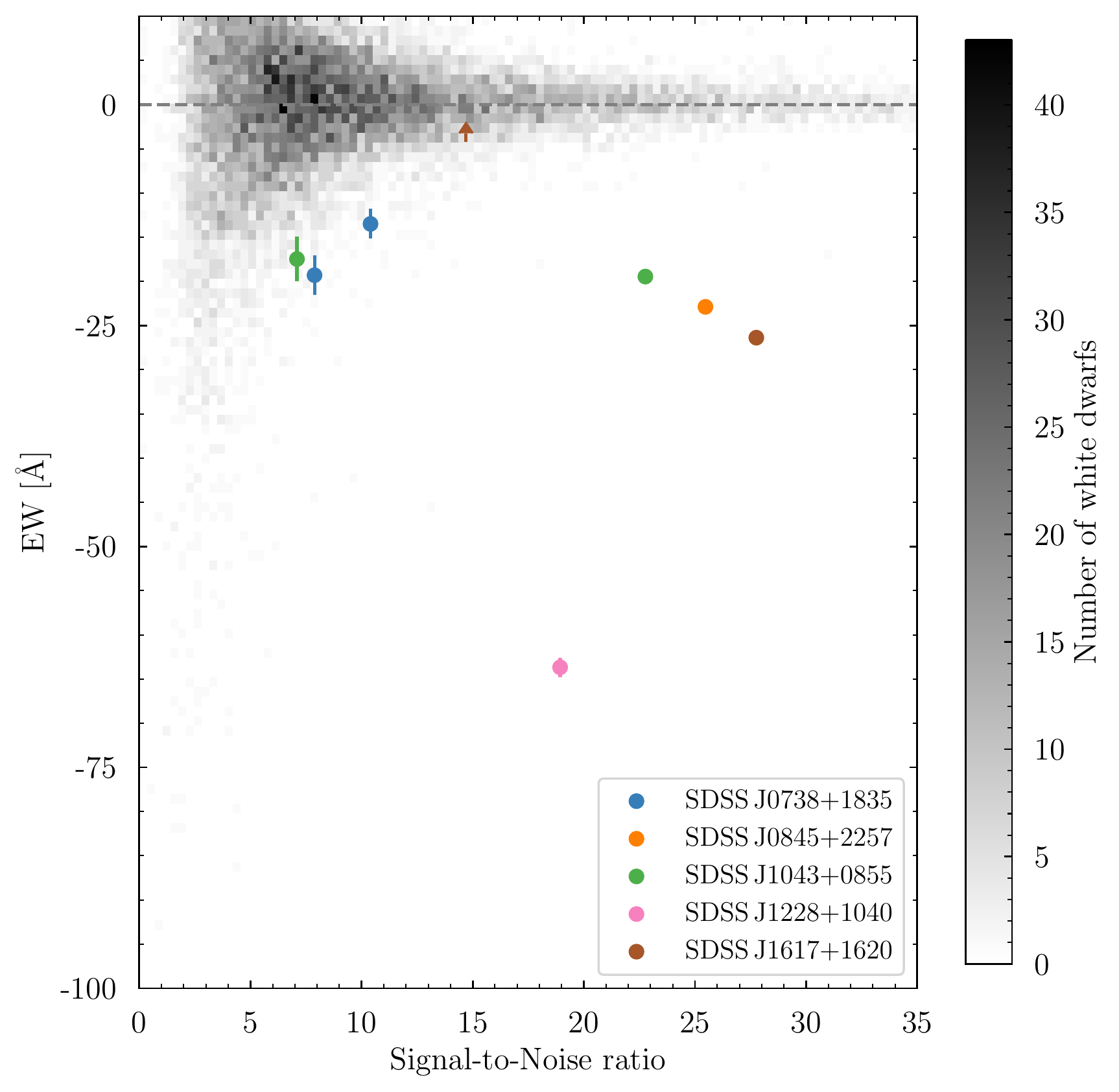}
\includegraphics[width=1\columnwidth]{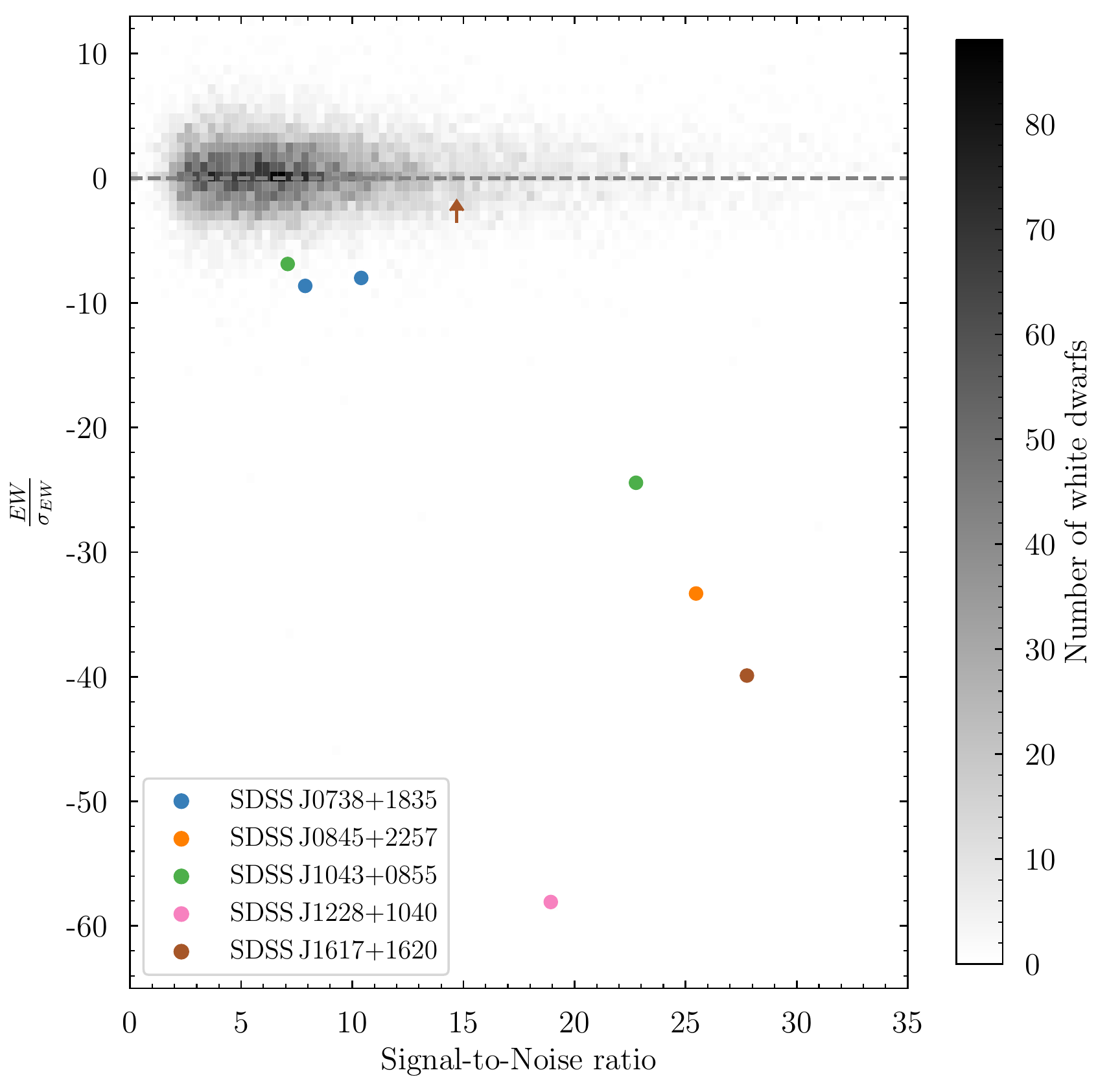}}
\caption{\label{f-StN} Left: A 2D density histogram for the Equivalent Width (EW) of the spectra (including reobservations) from our \textit{Gaia}/SDSS sample of 9374 single white dwarfs, with a signal-to-noise ratio (S/N) in the Ca\,{\textsc{ii}} region, S/N\,$\leq$\,35. The dashed line lies at zero, and observations of the five known gaseous discs in the sample are marked by the coloured circles. Arrows are upper limits on the EW in the case of non-detections. Some error bars are smaller than the symbols. Right: Similar to the left panel, but showing the distribution of $\frac{\textrm{EW}}{\sigma_{\textrm{EW}}}$. All spectra were visually inspected for the presence of Ca\,{\textsc{ii}} triplet emission.}
\end{figure*} 

We probed for the presence of gaseous emission by calculating the equivalent width (EW, a negative value corresponds to an emission feature) of the Ca\,{\textsc{ii}} triplet region between 8450\,\AA\ and 8700\,\AA, selecting spectra that satisfy $\frac{-\textrm{EW}}{\sigma_{\textrm{EW}}}$\,$\geq$\,3, where $\sigma_{\textrm{EW}}$ is the uncertainty on the EW measurement. While this method allows us to quantitatively inspect the sample, it is subject to poor sky subtraction and other observational artefacts which can contaminate the Ca\,{\textsc{ii}} triplet region. As such, we also visually inspected the continuum normalised Ca\,{\textsc{ii}} triplet 8600\,\AA\ region of all 14\,040 spectra in our sample to probe for the presence of emission. An example spectrum is shown in Figure\,\ref{f-normplot} with the Ca\,{\textsc{ii}} triplet region highlighted as an inset. After inspecting the entire spectroscopic sample we recovered the five known gaseous debris disc hosts in our sample\footnote[2]{The white dwarf SDSS\,J1144+0529 is also present in our sample which was identified by \cite{guoetal15-1} as hosting a gaseous debris disc. However, follow-up spectroscopy shows the emission lines to be radial-velocity variable, implying that it is a short-period binary containing a white dwarf and a low-mass companion (Florez et al. in prep), and we therefore exclude this system from the further discussion.}: SDSS\,J073842.57+183509.6, SDSS\,J084539.17+225728.0, SDSS\,J104341.53+085558.2, SDSS\,J122859.93+104032.9, and SDSS\,J161717.04+162022.4 (henceforth SDSS\,J0738+1835, SDSS\,J0845+2257, SDSS\,J1043+0855, SDSS\,J1228+1040, and SDSS\,J1617+1620 respectively), but did not find any new systems (see Figure\,\ref{f-normplot_appendix} for all eight spectra of the five gaseous debris disc systems). 

Figure\,\ref{f-StN} shows the distribution of EW and $\frac{\textrm{EW}}{\sigma_{\textrm{EW}}}$ against the signal-to-noise ratio (S/N) of the Ca\,{\textsc{ii}} triplet region for our white dwarf sample with S/N\,$\leq$\,35. Roughly 38\,per\,cent of the white dwarfs in our \textit{Gaia}/SDSS sample have multiple spectra, and repeat observations are included in Figure\,\ref{f-StN}. Measurements from spectra with detected gaseous emission are shown as filled colour circles. The non-detection of gaseous emission in one of the two spectra available for SDSS\,J1617+1620 is indicated by the arrow, and was the first of the two spectra obtained for this object \footnote[3]{SDSS\,J1617+1620 is the only system that shows a dramatic change in the EW of the Ca\,{\textsc{ii}} triplet where the SDSS spectra reveals a large increase in the EW. Subsequent follow-up observations recorded a gradual decrease in the EW of the Ca\,{\textsc{ii}} profile over 8\,yr (for more details see \citealt{wilsonetal14-1}). Variations in the morphology of the Ca\,{\textsc{ii}} triplet feature are seen in all other gaseous debris discs with long-term monitoring and is discussed later.}. Four gaseous disc observations are clearly separated from the bulk of the white dwarf spectra: SDSS\,J1228+1040, SDSS\,J1043+0855, SDSS\,J0845+2257; the first three of such discs discovered \citep{gaensickeetal06-3, gaensickeetal07-1, gaensickeetal08-1}, and SDSS\,J1617+1620 \citep{wilsonetal14-1}.

The spread in EW values is extremely large at low S/N values due to noise (compare the left and right panels of Figure\,\ref{f-StN}), but rapidly tightens as the S/N increases, making it trivial to identify genuine emission features such as SDSS\,J1228+1040. Closer to the bulk EW distribution, it becomes difficult to detect Ca\,{\textsc{ii}} triplet emission such as that of SDSS\,J0738+1835 and SDSS\,J1043+0855 (see Figure\,\ref{f-normplot_appendix} for the EW values). We therefore chose to limit the sample used for the following statistical analysis to spectra with S/N\,$\simeq$\,5, wherein if there were any additional emission features with an EW similar to the gaseous debris discs in our sample we should have detected them. Assuming that we can only detect gaseous emission in the spectra of systems above S/N\,$=$\,5, the number of white dwarfs in our sample reduces to 7705, which we use for the calculations in the next section.

\subsection{The occurrence rate of gaseous discs at white dwarfs}\label{s-occ-rate}

We find no new gaseous discs in our magnitude  and S/N limited \textit{Gaia}/SDSS sample, resulting in a total of five detected gaseous components to debris discs out of 7705 systems. Using Bayes' theorem we calculate the probability distribution, $p(f|n,t)$, of an occurrence rate, $f$, of a gaseous debris disc at a white dwarf, given $n$ detections out of $t$ systems as

\begin{equation}\label{e-bayes}
p(f|n,t) = \frac{p(n|f,t) p(f)}{p(n,t)},
\end{equation}

\noindent where $p(n|f,t) = f^{n}(1-f)^{t-n}$ is the binomial distribution, $p(f)$ is the prior distribution of $f$, and $p(n,t)$ is the probability distribution of detecting $n$ gaseous disc hosts out of $t$ systems which is a constant. We assume an uninformative prior distribution known as the Jeffreys prior \citep{jeffreys46-1}, which for the binomial distribution is

\begin{equation}
p(f) = \frac{f^{-0.5}(1-f)^{-0.5}}{B(0.5,0.5)},
\end{equation}

\noindent where $B$ is the beta function. Substituting this prior into Equation\,\ref{e-bayes} results in

\begin{equation}\label{e-final}
p(f|n,t) \propto f^{n-0.5}(1-f)^{t-n-0.5}.
\end{equation}

\begin{figure}
\centerline{\includegraphics[width=1\columnwidth]{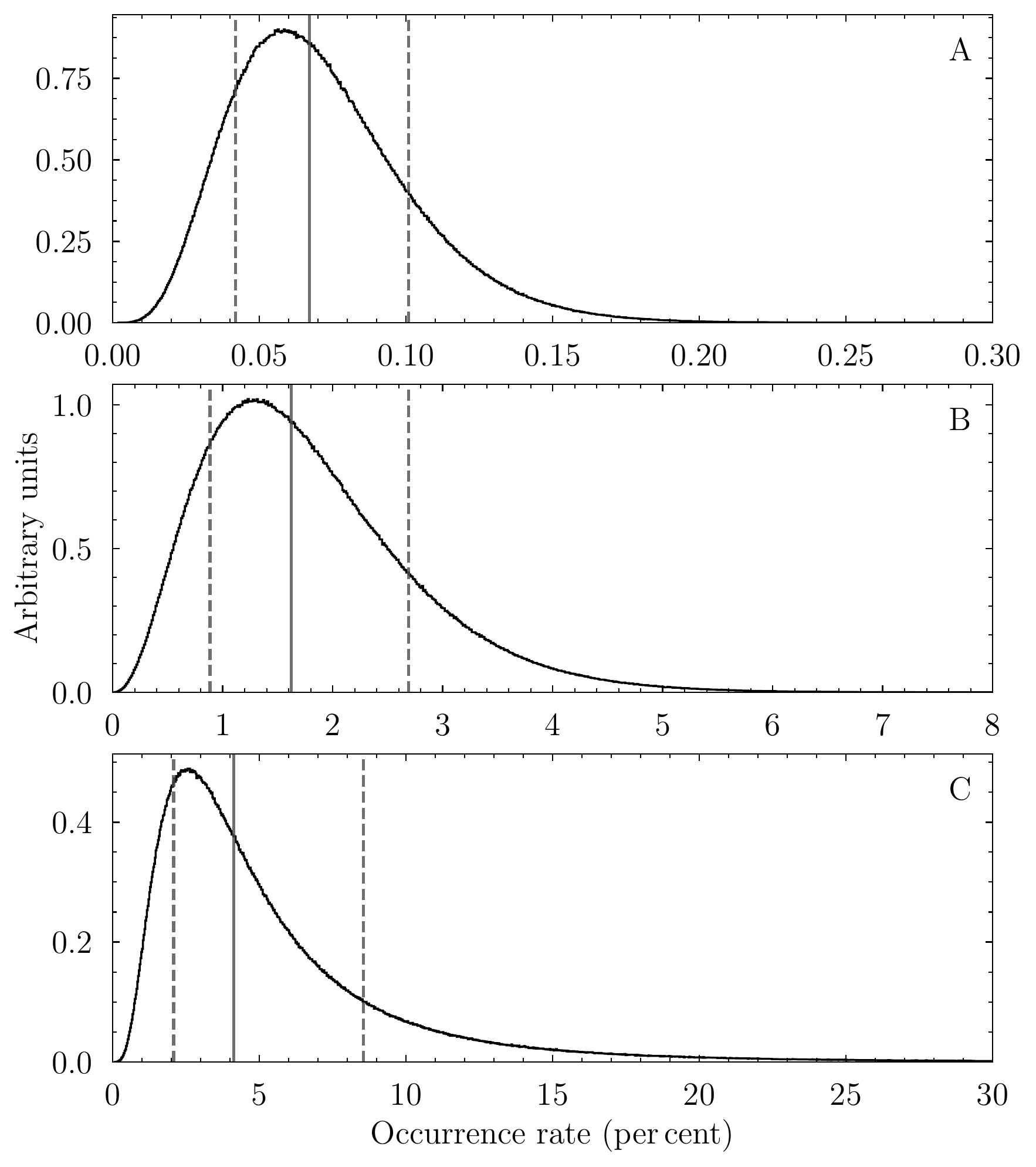}}
\caption{\label{f-frac} Histograms evaluated from the probability distributions of the occurrence rates of (A) debris discs with a gaseous component in emission around white dwarfs, (B) dusty debris discs around white dwarfs, and (C) a gaseous component to a debris disc in emission. The solid gray lines represent the median value of the distributions, and the dashed lines mark the 15.9 and 84.1 percentiles corresponding to $\pm$\,1\,$\sigma$. (A) [(B)] is constructed from 10 million evaluations of Equation\,\ref{e-final} using $n$ = 5 [3], and $t$ = 7705 [195]. (C) is produced by dividing each evaluation of (A) by a corresponding evaluation of (B).}
\end{figure}

\noindent Taking the median and 1$\sigma$ intervals of this probability distribution (which is independent of the proportionality constant) with $n=5$, and $t=7705$, we find that 0.067\,$\pm$\,$_{0.025}^{0.042}$\,per\,cent of white dwarfs host a debris disc with a detectable gaseous component (Figure\,\ref{f-frac}\,A). This is significantly lower than the occurrence rate of detectable dust discs around white dwarfs, which has been estimated by various studies to be 1--3\,per cent \citep{farihietal09-1, girvenetal11-1, rocchettoetal15-1, wilsonetal19-1, rebassa-mansergasetal19-1}. For further discussion, we adopt the estimate of \citet{wilsonetal19-1}, 1.5\,$\pm$\,$_{0.5}^{1.5}$\,per\,cent, calculated using an unbiased sample of 195 white dwarfs observed with \textit{Spitzer} containing three dusty debris discs. We note, however, that using Equation\,\ref{e-final} with $n=3$ and $t=195$, we obtain a slightly different occurrence rate of 1.6\,$\pm$\,$_{0.7}^{1.1}$\,per\,cent, and the probability distribution of this occurrence rate is shown in Figure\,\ref{f-frac}\,B. These two occurrence rates of dusty debris discs at white dwarfs are in agreement, and the difference results from the choice of the prior distribution. We use the result calculated in this manuscript in our next steps for internal consistency within our analysis. If we assume that systems with a gaseous component in emission are a subset of the systems that host dusty debris, then the ratio of the two distributions representing the dusty (Figure\,\ref{f-frac}\,A) and the gaseous (Figure\,\ref{f-frac}\,B) provides the probability distribution for the occurrence rate of a debris disc to host a gaseous component in emission (Figure\,\ref{f-frac}\,C). By taking the median value and 1$\sigma$ intervals of the distribution in Figure\,\ref{f-frac}\,C, we determine that 4\,$\pm$\,$_{2}^{4}$\,per\,cent of dusty debris discs should also host a detectable gaseous component. This value is a factor of a few smaller than the value of $\simeq$\,19\,per cent, which is obtained from simply taking the ratio of known gas (7) to dust ($\simeq$\,37) discs around metal polluted white dwarfs. This is due to the fact that all five stars in our sample that have Ca\,{\textsc{ii}} triplet emission were discovered from their SDSS spectra, and their infrared emission was only detected \textit{after} the identification of the gaseous discs in targeted, deep \textit{Spitzer} observations \citep{brinkworthetal09-1, dufouretal10-1, brinkworthetal12-1}.

\section{Observations of debris discs}
\label{sec:observations}

An independent assessment of the fraction of dusty debris discs also having a gaseous component can be obtained from inspecting the known sample of white dwarfs with circumstellar dust discs detected by an infrared excess, and corroborated by the detection of photospheric metal contamination. This requires red spectroscopy covering the Ca\,{\textsc{ii}} 8600\,\AA\ triplet region, probing for emission lines. Table\,\ref{t-logwht} lists 37 white dwarf systems where a dusty debris disc has been confirmed. For completeness, we also list five candidate debris disc systems, however we do not include those in the statistical analysis. This sample includes five white dwarfs that were observed by \textit{Spitzer} only because they were identified to have Ca\,{\textsc{ii}} triplet emission (see Table\,\ref{t-logwht}); a clear indicator for the presence of a disc. Without the detection of a gaseous disc, based on optical spectroscopy, these systems would have not been observed with \textit{Spitzer} (i.e. these observations were obtained \textit{after already knowing} that these systems harbour a debris disc). These observations are clearly biased, and we therefore remove these five gaseous debris discs from our sample, reducing the number of systems included to 32. White dwarfs being blue objects with, usually, no features of interest red-ward of H$\alpha$, it is unsurprising that only 17 of the remaining 32 confirmed dusty debris discs around metal polluted white dwarfs have published spectroscopy covering the Ca\,{\textsc{ii}} triplet region. 

\begin{table*}
\centering
\caption{White dwarfs with infrared excesses attributed to dusty debris discs. Observations presented in this work are detailed with the telescope and instrument used, date, and exposure time. The five systems in the bottom part of the table display an infrared excess, however, the presence of a dusty disc remains so far unconfirmed and are therefore not used in our analysis. \label{t-logwht}}
\begin{tabular}{llllrlr}
\hline
WD & Alternate name & Telescope/Instrument & Date & Exposure time [s] & Gaseous emission & Ref. \\
\hline
0010+280  & PG\,0010+281 &   &   &  & No & 1, 2, 3\\
0106--328$^{\textrm{a}}$  & HE\,0106--3253 & SOAR/GHTS & 2017\,-\,08\,-\,26 & 2400& No & 4, 5 \\
0110--565$^{\textrm{a}}$  & HE\,0110--5630 & SOAR/GHTS & 2017\,-\,08\,-\,26 & 2400& No & 5, 6 \\
0146+187$^{\textrm{a}}$  & GD\,16 & WHT/ISIS & 2011\,-\,12\,-\,07 & 4500& No & 7, 8 \\
0300--013$^{\textrm{a}}$ & GD\,40 & WHT/ISIS & 2011\,-\,12\,-\,07 & 6600& No & 5, 9\\
0307+077 & HS\,0307+0746 & WHT/ISIS & 2011\,-\,12\,-\,06 & 4500& No & 4\\
0408--041 & GD\,56 & WHT/ISIS & 2011\,-\,12\,-\,07 & 4200& No & 9\\
0420+520 & KPD\,0420+5203 &   &   & & Unknown & 1 \\
0435+410 & GD\,61 & WHT/ISIS & 2011\,-\,12\,-\,07 & 2700& No & 6\\
0735+187  & SDSS\,J0738+1835 &   &   &  & Yes$^{\textrm{b}}$ & 10, 11 \\
0842+231  & SDSS\,J0845+2257 &   &   &  & Yes$^{\textrm{b}}$ & 12, 13 \\
- & EC\,05365--4749 & SOAR/GHTS & 2017\,-\,08\,-\,26 & 3000& No & 14\\
0843+516 & PG\,0843+517 & WHT/ISIS & 2011\,-\,12\,-\,06 & 3600& No & 15\\
0956--017$^{\textrm{c}}$  & SDSS\,J0959-0200 & & & & Yes & 16, 17 \\
1015+161 & PG\,1015+161 & WHT/ISIS & 2011\,-\,12\,-\,08 & 3000 & No & 9\\
1018+410  & PG\,1018+411 & & &  & No & 17, 18\\
1041+091  & SDSS\,J1043+0855 & & & & Yes$^{\textrm{b}}$ & 13, 19\\
1116+026 & GD\,133 & WHT/ISIS & 2011\,-\,12\,-\,07 & 2700& No & 9\\
1145+017 & LBQS\,1145+0145 & & && No & 20, 21\\
1150--153 & EC\,11507--1519 & WHT/ISIS & 2011\,-\,12\,-\,07 & 3600& No & 22\\
1219+130  & SDSS\,J1219+1244 & & && No & 16, 23 \\
1225--079 & PG\,1225--079 & & & & No & 3, 24 \\
1226+110  & SDSS\,J1228+1040 & & & & Yes$^{\textrm{b}}$ & 25, 26 \\
1349--230 & HE\,1349--2305 & & & & Yes & 5, 27 \\
1455+298 & G\,166-58 & WHT/ISIS & 2017\,-\,06\,-\,28 & 1800 & No & 28\\
1457--086 & PG\,1457--086 & WHT/ISIS & 2010\,-\,07\,-\,24 & 2700 & No & 7\\
1541+650 & PG\,1541+651 & WHT/ISIS & 2017\,-\,04\,-\,02 & 1800 & No & 29 \\
1551+175 & KUV\,15519+1730 & & & & No & 18, 30\\
1554+094 & SDSS\,J1557+0916 & & & & No & 14, 23\\
1615+164 & SDSS\,J1617+1620 & & & & Yes$^{\textrm{b}}$ & 10, 13\\
1729+371$^{\textrm{a}}$ & GD\,362 & WHT/ISIS & 2006\,-\,07\,-\,01 & 900 & No & 31 \\
1929+011  & GALEX\,J1931+0117 & & & & No & 32, 33 \\
2115--560 & LAWD\,84 & & & & No & 6, 34\\
2132+096 & HS\,2132+0941 & WHT/ISIS & 2016\,-\,10\,-\,29 & 1800 & No & 32\\
2207+121  & SDSS\,J2209+1223 & & & & No & 14, 18\\
2221--165 & HE\,2221--1630 & WHT/ISIS & 2017\,-\,06\,-\,28 & 1800 & No & 4\\
2326+049 & G\,29-38 & WHT/ISIS & 2016\,-\,10\,-\,29 & 1800 & No & 35\\
\hline
0145+234 & Mrk\,362 & & & & Unknown & 36 \\
- & LSPM\,J0207+3331 & & & & Unknown & 30 \\
0246+734 & EGGR\,474 & & & & No & 32 \\
0950--572 & NLTT\,22825 &   &   & & Unknown & 1 \\
2328+107 & PG\,2328+108 & WHT/ISIS & 2016\,-\,10\,-\,29 & 1800 & No & 37\\
\hline
\end{tabular}
\begin{minipage}{15cm}
 \small
$^{\textrm{a}}$ These systems already have published spectra covering the Ca\,{\textsc{ii}} triplet region.\\
$^{\textrm{b}}$ Gaseous Ca\,{\textsc{ii}} triplet emission from the discs in our sample were identified prior to the detection of an infrared excess.\\
$^{\textrm{c}}$ System falls in the SDSS footprint, but does not have SDSS spectroscopy.\\
\textbf{References:}
(1) \cite{barberetal16-1}
(2) \cite{xuetal15-1}
(3) \cite{xuetal19-2}
(4) \cite{farihietal10-1}
(5) \cite{dennihyetal17-1}
(6) \cite{girvenetal12-1}
(7) \cite{farihietal09-1}
(8) \cite{fusilloetal17-1}
(9) \cite{juraetal07-1} 
(10) \cite{gaensicke11-1}
(11) \cite{dufouretal10-1}
(12) \cite{gaensickeetal08-1}
(13) \cite{brinkworthetal12-1}
(14) \cite{dennihyetal16-1}
(15) \cite{xu+jura12-1}
(16) \cite{girvenetal11-1}
(17) \cite{farihietal12-1}
(18) \cite{fusilloetal15-1}
(19) \cite{gaensickeetal07-1}
(20) \cite{vanderburgetal15-1}
(21) \cite{xuetal16-1}
(22) \cite{juraetal09-1}
(23) \cite{zuckermanetal03-1}
(24) \cite{kleinetal11-1}
(25) \cite{gaensickeetal06-3}
(26) \cite{brinkworthetal09-1}
(27) \cite{melisetal12-1}
(28) \cite{farihietal08-1}
(29) \cite{barberetal12-1}
(30) \cite{debesetal19-1}
(31) \cite{juraetal07-2}
(32) \cite{bergforsetal14-1} 
(33) \cite{vennesetal10-1}
(34) \cite{swanetal19-1}
(35) \cite{zuckerman+becklin87-1}
(36) \cite{wangetal19-1}
(37) \cite{rocchettoetal15-1}
\end{minipage}
\end{table*}

\begin{figure}
\centerline{\includegraphics[width=1\columnwidth]{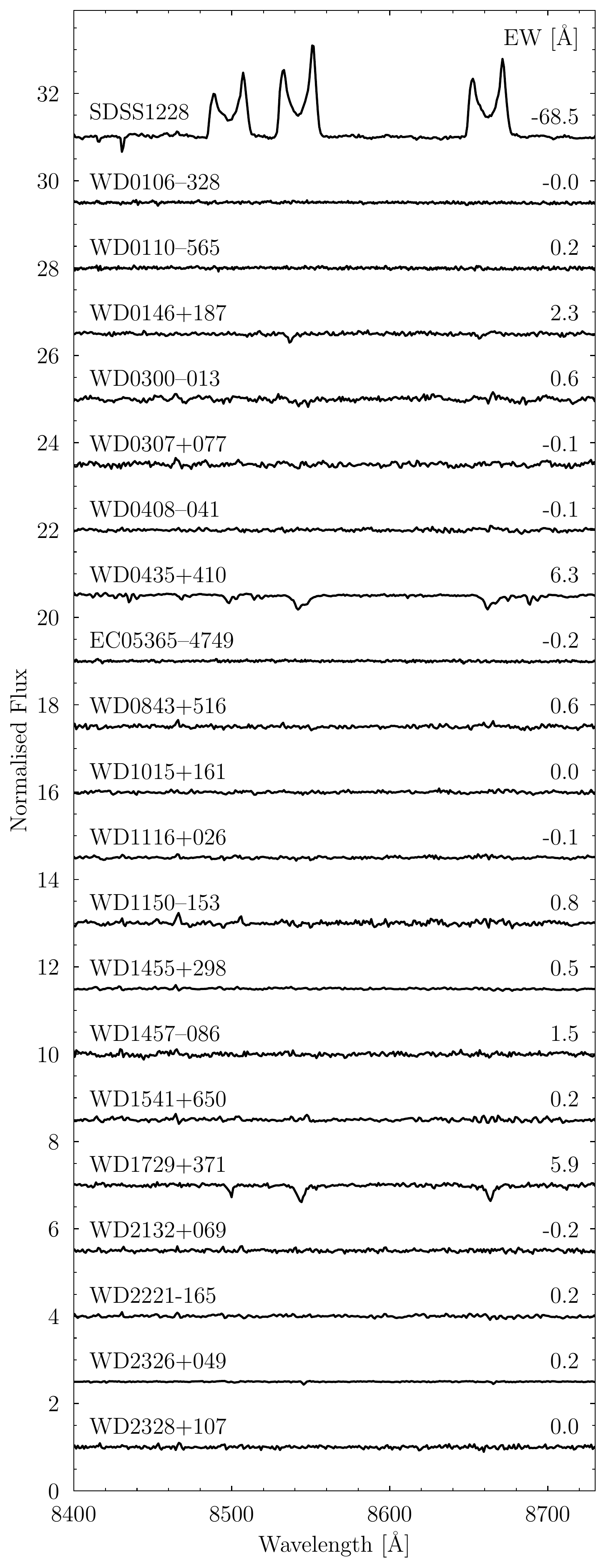}}
\caption{\label{f-dusty} The continuum-normalised Ca\,{\textsc{ii}} triplet region of 20 white dwarfs with a detected infrared excess attributed to a dusty disc. The emission profile of SDSS\,J122859.93+104032.9 (SDSS1228) is included for comparison, and the equivalent widths for each profile are given. The normalised spectra are shifted in steps of 1.5 units with respect to the spectrum of WD\,2328+107.}
\end{figure}

To rectify this, we obtained spectroscopy of 20 white dwarfs with an infrared excess (five of which had been previously observed) using the Intermediate dispersion Spectrograph and Imaging System (ISIS) on the William Herschel telescope (WHT), and the Goodman High Throughput Spectrograph (GHTS) on the Southern Astrophysical Research Telescope (SOAR). The details of the observations are listed in Table\,\ref{t-logwht}. The WHT spectra were reduced using standard spectroscopic techniques, i.e. bias-subtraction, flat-fielding, sky-subtraction, and optimal-extraction of the 1D spectra. These were achieved using the \textsc{kappa}, \textsc{figaro}, and \textsc{pamela} packages within the \textsc{starlink} software distribution. The resulting 1D spectra were then wavelength calibrated (CuNe+CuAr arc-lamps) and flux calibrated using \textsc{molly}\footnote[4]{Molly software can be found at \url{http://deneb.astro.warwick.ac.uk/phsaap/software/molly/html/INDEX.html}}. SOAR spectroscopic data were reduced using iraf's noao package. The frames were first bias-subtracted, and flattened with a quartz lamp flat. We subtracted the background and extracted the spectra with the apall task, and finally performed wavelength calibration with a CuHeAr lamp spectrum extracted with the same aperture. The normalised spectra of the Ca\,{\textsc{ii}} region are shown in Figure\,\ref{f-dusty}. None of the 20 white dwarfs we observed show the characteristic Ca\,{\textsc{ii}} emission profiles in the 8600\,\AA\ region. We note that WD\,0146+187, WD\,0435+410, WD\,1729+371, and WD\,2326+049 do show photospheric Ca\,{\textsc{ii}} triplet absorption due the metal pollution of the white dwarf atmosphere.

Including our observations WD\,0420+520 is the only confirmed planetary debris disc system without a published spectrum of the Ca\,{\textsc{ii}} triplet, and we therefore remove it from our sample leaving 31 systems. Our sample contains white dwarfs that have both (i) been initially identified as having a dusty disc, and (ii) follow-up spectroscopy of the Ca\,{\textsc{ii}} triplet region, which we use to calculate an independent measure of the occurrence rate of observable gaseous components to debris discs (see Table\,\ref{t-logwht}). The sample contains two gaseous debris discs; SDSS\,J0959--0200 and HE\,1349--2305, and using Eqn.\,\ref{e-final} with $n = 2$ and $t = 31$, we determine an occurrence rate of a detectable gaseous component in emission among dusty debris discs as 7\,$\pm$\,$_{4}^{5}$\,per\,cent which it is consistent with the occurrence rate we determined in Section\,\ref{s-occ-rate} of 4\,$\pm$\,$_{2}^{4}$\,per\,cent.

\section{Discussion}
\label{sec:discussion}

The majority of gaseous components to debris discs with long-term monitoring show similar morphological variations of their line profiles on time-scales of years to decades \citep{wilsonetal15-1, manseretal16-1, manseretal16-2, dennihyetal18-1}. The line profile variations of the gaseous disc at SDSS\,J1228+1040 have been interpreted as arising from a fixed, eccentric intensity pattern that slowly precesses with a period of $\simeq$\,27\,yr \citep{manseretal16-1}. Short cadence observations of the Ca\,{\textsc{ii}} triplet at SDSS\,J1228+1040 also show 3\,--\,4\,per\,cent variability in the strength of emission on a $\simeq$\,2\,hr period generated by a planetesimal with high internal strength orbiting within the Roche radius \citep{manseretal19-1}. In contrast, the emission from the disc at HE\,1349--2305 precesses on a significantly shorter time-scale of $\simeq$\,1.4\,yr \citep{dennihyetal18-1}. The Ca\,{\textsc{ii}} triplet emission from SDSS\,J0845+2257 and SDSS\,J1043+0855 show similar morphological variations on a roughly decadal time-scale thought to be produced by apsidal precession, although they have not yet been shown to be periodic \citep{gaensickeetal08-1, wilsonetal15-1, manseretal16-2}. 

The order-of-magnitude difference between the time-scales of variability seen at these systems is not yet understood, but their similar behaviour points to a common underlying mechanism. The gas emission profiles allow constraints to be placed on the radial extent of the gaseous debris, which show that they are all co-orbital with the dusty debris in the disc \citep{brinkworthetal09-1,melisetal10-1}. 

Several models have been proposed to explain the presence of a gaseous component to debris discs that is co-orbital with the dusty component. \cite{rafikov11-2} and \cite{metzgeretal12-1} proposed that the production of gas at the sublimation radius will lead to gas spreading both inwards towards the white dwarf and outwards into the debris disc due to angular momentum exchange. The outward moving gas will cause the dust particles in the disc to experience aerodynamic drag, which enhances the rate at which dust particles are fed past the sublimation radius. This leads to a runaway process where large amounts of gas are produced and the accretion rate onto the white dwarf increases by orders of magnitude. However, an important problem with this model is that gas which is co-orbital with dust should condense back into dust and therefore be depleted over several hundred orbital time-scales \citep{metzgeretal12-1}. Consequently, without a mechanism that maintains sufficient amounts of material streaming radially outwards in the gaseous phase, a runaway process cannot be achieved.

\cite{kenyon+bromley17-2} outline a model where a collisional cascade of solid material vaporised within the Roche limit of a white dwarf can feed a gaseous component of a debris disc (see also \citealt{kenyon+bromley17-1}). This model can explain the presence of gaseous material co-orbital with dusty debris, in addition to the accretion rates observed at white dwarfs with debris discs, but it cannot elucidate why we only detect gas in a subset of debris discs.

No process has been proposed that explains the combined rarity, eccentricity, and (periodic in two cases) variability seen in most of the gaseous components of debris discs. We hypothesise here that these components are generated by collisions induced by a planetesimal on a close-in eccentric orbit around the white dwarf \citep{jura08-1, farihietal09-1, manseretal19-1}. It is possible that the planetesimal is also perturbing dust out of a flat disc geometry, where the dust is no longer shielded within the disc, is subject to sublimation by direct irradiation from the white dwarf \citep{rafikov+garmilla12-1}. This process could also lead to variations in the infrared luminosity of the dusty component of debris discs around white dwarfs through dust production and destruction (e.g. see \citealt{farihietal18-1} for an in-depth discussion). Such infrared luminosity variations have been seen at a number of debris discs around white dwarfs \citep{xu+jura14-1, farihietal18-1, swanetal19-2, wangetal19-1}, including SDSS\,J1228+1040 \citep{xuetal18-1}.

In our qualitative model, the planetesimal produces gas that initially traces the orbit of the planetesimal. This gas is built up, and begins to circularise and to spread radially, forming the disc detected in the metallic emission lines (Figure\,\ref{f-normplot}). Eventually this gas condenses back into dust. This requires that the condensation time-scale is (1) similar to the circularisation time-scale, and (2) significantly shorter than the viscous spreading time-scale. Whereas the viscous time-scale is thought to be on the order of years to decades \citep{shakura+sunyaev73-1,rafikov11-1, wilsonetal14-1}, further work is needed to constrain the circularisation time-scale for these discs.

The framework outlined above explains observational features of gaseous debris discs and remnant planetary systems around white dwarfs:

(i) The rarity of gaseous debris discs could be explained by the presence of a planetesimal that is actively disrupting the disc. The occurrence rate we calculate in this paper would imply directly a lower limit on the occurrence rate of planetesimals orbiting within debris discs around white dwarfs. In debris discs without a planetesimal generating a significant source of gaseous material, any gas should condense into dust and no gaseous component would be detected.

(ii) There is no observed correlation between the presence of a gaseous component to a debris disc and the accretion rate of metals onto the disc-hosting white dwarf (see fig.\,10 of \citealt{farihi16-1} and table 2 of \citealt{manseretal16-2}). This is easily explained by the gaseous component of the disc recondensing into dust before it has enough time to viscously spread and accrete onto the white dwarf.

(iii) \cite{manseretal19-1} show in their fig.\,S2 that the EW of the Ca\,{\textsc{ii}} triplet profiles can vary as much as 20\,per\,cent over a few weeks to months. This time-scale is equivalent to a few hundred orbital time-scales (similar to the estimates of the condensation time-scale, \citealt{metzgeretal12-1}), and could be explained by a change in the rate at which the planetesimal is generating gas through dust destruction and disc warping leading to sublimation. The majority of Ca\,{\textsc{ii}} triplet observations were obtained on yearly time-scales, and more frequent monitoring of the Ca\,{\textsc{ii}} triplet emission profiles and their EW is required to establish in detail the extent of this weekly to monthly variability.

(iv) The observed eccentricity and precession of the gaseous debris discs \citep{manseretal16-1, dennihyetal18-1} could be induced by a planetesimal on an eccentric orbit. \cite{miranda+rafikov18-1} show that an eccentric gas disc will precess on a fixed period, but suggest that a method to excite the eccentricity must exist in these systems. In our scenario, the precession of the gaseous debris disc component naturally arises from the effect of general relativity on the eccentric orbit of the planetesimal \citep{manseretal19-1}. The precession periods determined in SDSS\,J1228+1040 and HE\,1349--2305 differing by an order of magnitude would then reflect different orbital configurations of the planetesimals in these systems. One way to probe for the eccentricity and precession of the orbit of a planetesimal is to search for changes in the radial velocity trail of the gas it generates (see fig. 1\,C\,\&\,F of \cite{manseretal19-1}). Whereas the radial velocity variation of a test mass on a circular orbit will follow a sine wave, that of an eccentric orbit depends on the eccentricity of the orbit as well as the angle, $\beta$, between the semi-major axis of the orbit and the line of sight from the observer to the system. If the planetesimal is apsidally precessing due to general relativity, then $\beta$ will vary periodically with the precession period, which is an observationally verifyable prediction of our hypothesis.

\section{Conclusions}
\label{sec:conc}

We have determined the occurrence rate of gaseous debris discs in emission around white dwarfs as 0.067\,$\pm$\,$_{0.025}^{0.042}$\,per\,cent using a magnitude and S/N limited sample of white dwarfs observed by SDSS and \textit{Gaia}. This rate is significantly smaller than the ratio of observed debris discs with and without a gaseous component in emission due to different observational biases in the detection of emission lines from the gas and broad-band infrared emission from the dust. We also present the non-detection of Ca\,{\textsc{ii}} triplet emission at 20 dusty debris disc hosts. The forthcoming dedicated multi-object spectroscopic surveys (DESI, \citealt{DESI16-1, DESI16-2}, WEAVE, \citealt{daltonetal12-1, daltonetal16-1}, 4MOST, \citealt{dejongetal16-1}, and SDSS-V, \citealt{kollmeieretal17-1}) are expected to increase the number of known gaseous discs by a factor $\simeq$\,5 over the next decade. This much enlarged sample will provide detailed insight into the orbital properties of planetesimals closely orbiting white dwarfs, and into the mechanisms generating the detected gas.

\section*{Acknowledgements}
We thank the anonymous referee for their useful comments and meticulous reading of the manuscript. C.J.M. and B.T.G. were supported by the UK STFC grant ST/P000495. I.P. was supported by Capes-Brazil under grant 88881.134990/2016-01. Based on observations obtained at the Southern Astrophysical Research (SOAR) telescope, which is a joint project of the Minist\'{e}rio da Ci\^{e}ncia, Tecnologia, Inova\c{c}\~{o}es e Comunica\c{c}\~{o}es (MCTIC) do Brasil, the U.S. National Optical Astronomy Observatory (NOAO), the University of North Carolina at Chapel Hill (UNC), and Michigan State University (MSU). This work has made use of observations from the SDSS-III, funding for which has been provided by the Alfred P. Sloan Foundation, the Participating Institutions, the National Science Foundation, and the U.S. Department of Energy Office of Science. The SDSS-III web site is http://www.sdss3.org/. Based on observations made with the WHT operated on the island of La Palma by the Isaac Newton Group in the Spanish Observatorio del Roque de los Muchachos of the Instituto de Astrof\'isica de Canarias.

\bibliographystyle{mnras}
\bibliography{aamnem99,aabib}

\appendix
\newpage
\onecolumn

\section{SDSS spectroscopy of gaseous debris discs}\label{appendix-1}

\begin{figure}
\centerline{\includegraphics[width=1\columnwidth]{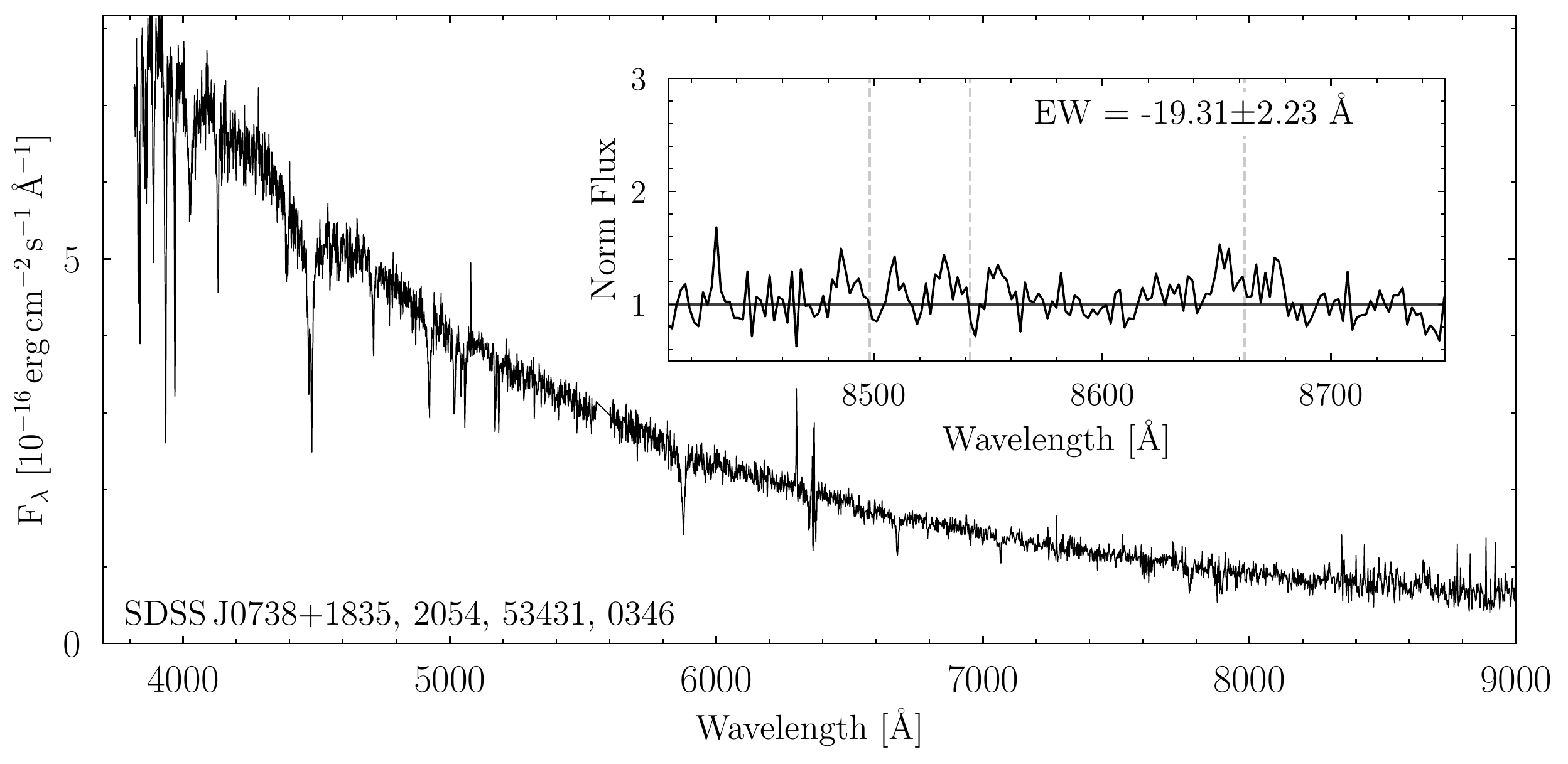}}
\centerline{\includegraphics[width=1\columnwidth]{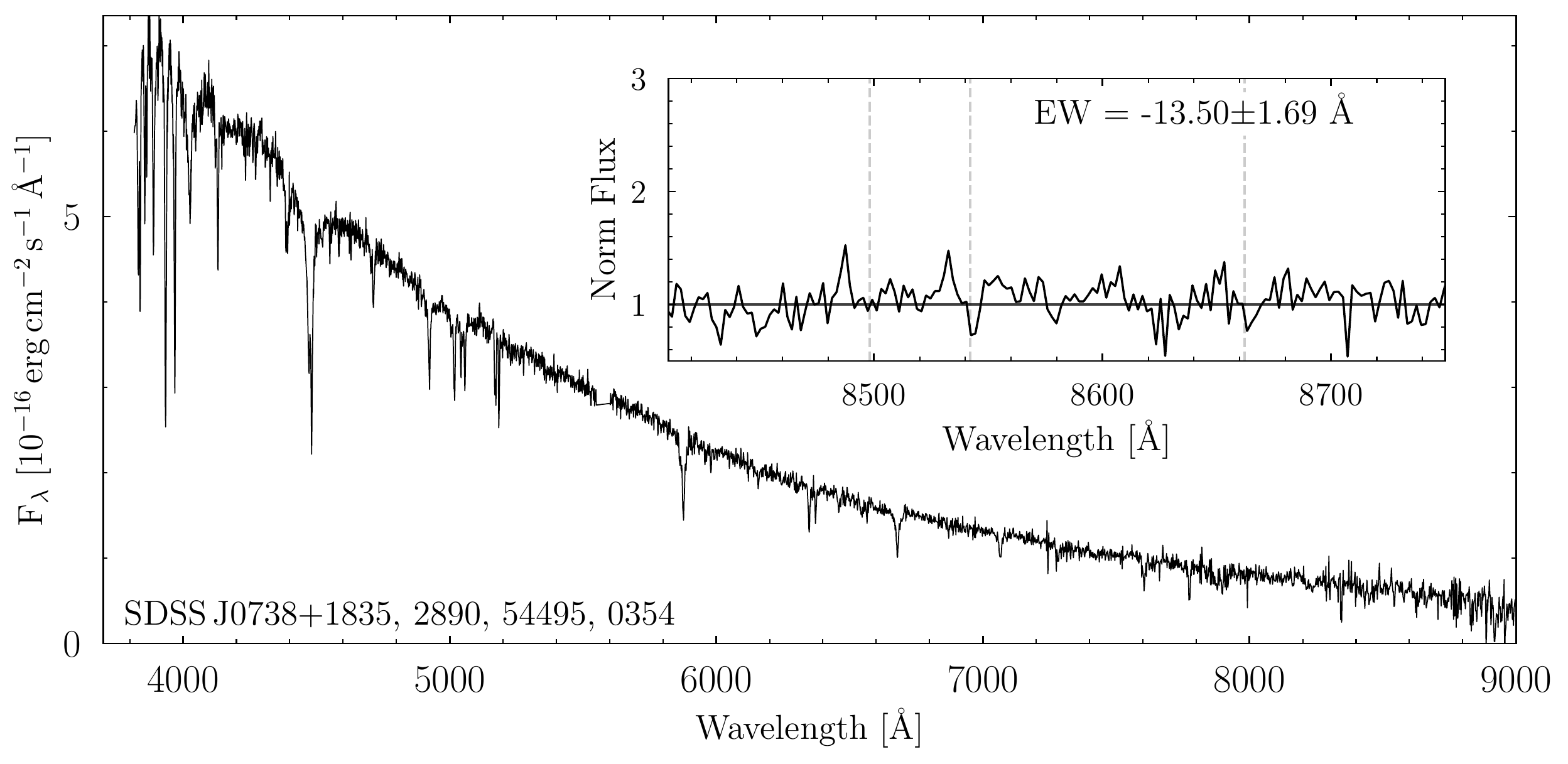}}
\caption{\label{f-normplot_appendix} Similar to Figure\,\ref{f-normplot} showing the complete set of SDSS spectra for the gaseous debris discs identified in our \textit{Gaia}/SDSS sample. The name, along with the SDSS Plate, MJD, and fiber identifiers are also given for each spectrum.}
\end{figure} 

\renewcommand{\thefigure}{A\arabic{figure}}
\setcounter{figure}{0}

\begin{figure}
\centerline{\includegraphics[width=1\columnwidth]{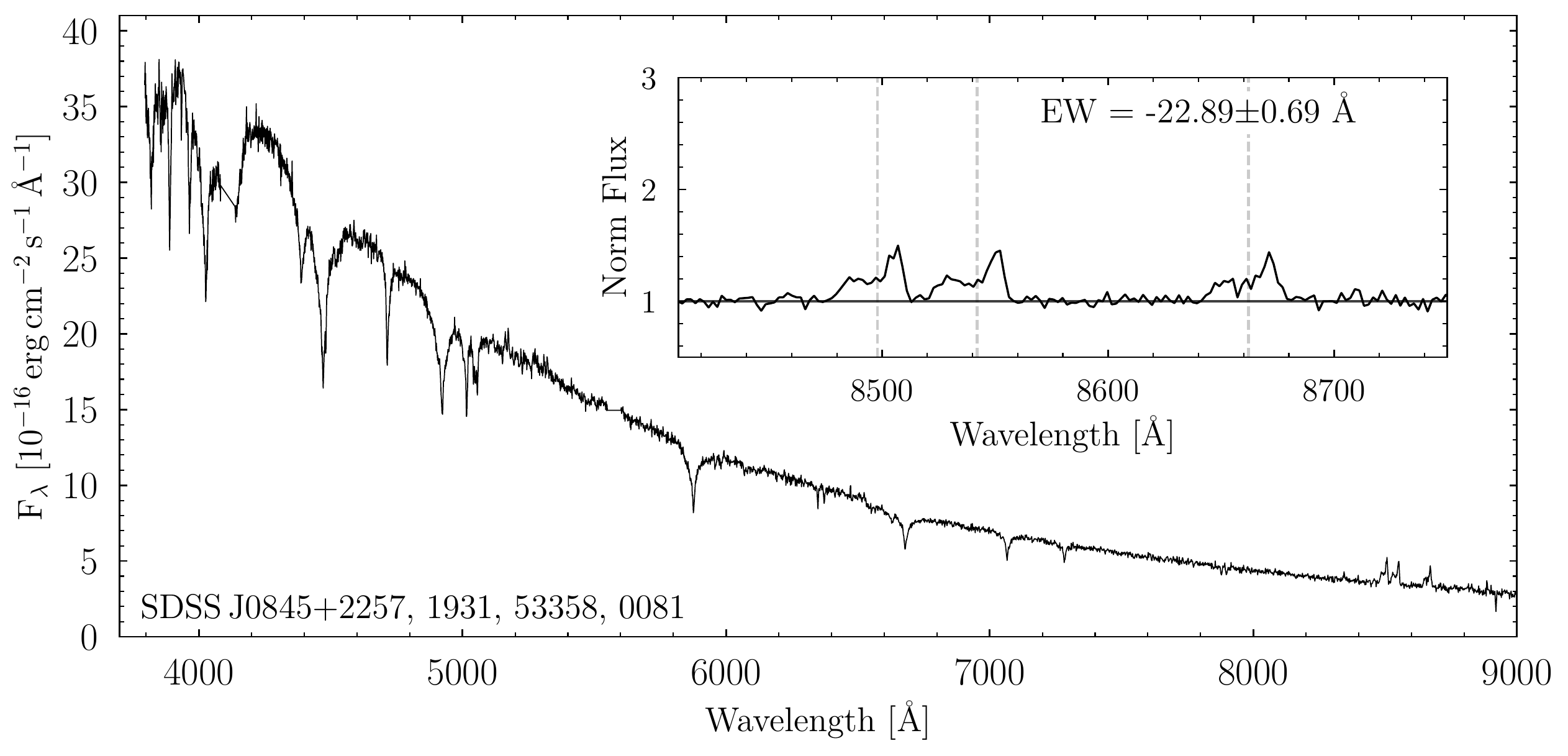}}
\centerline{\includegraphics[width=1\columnwidth]{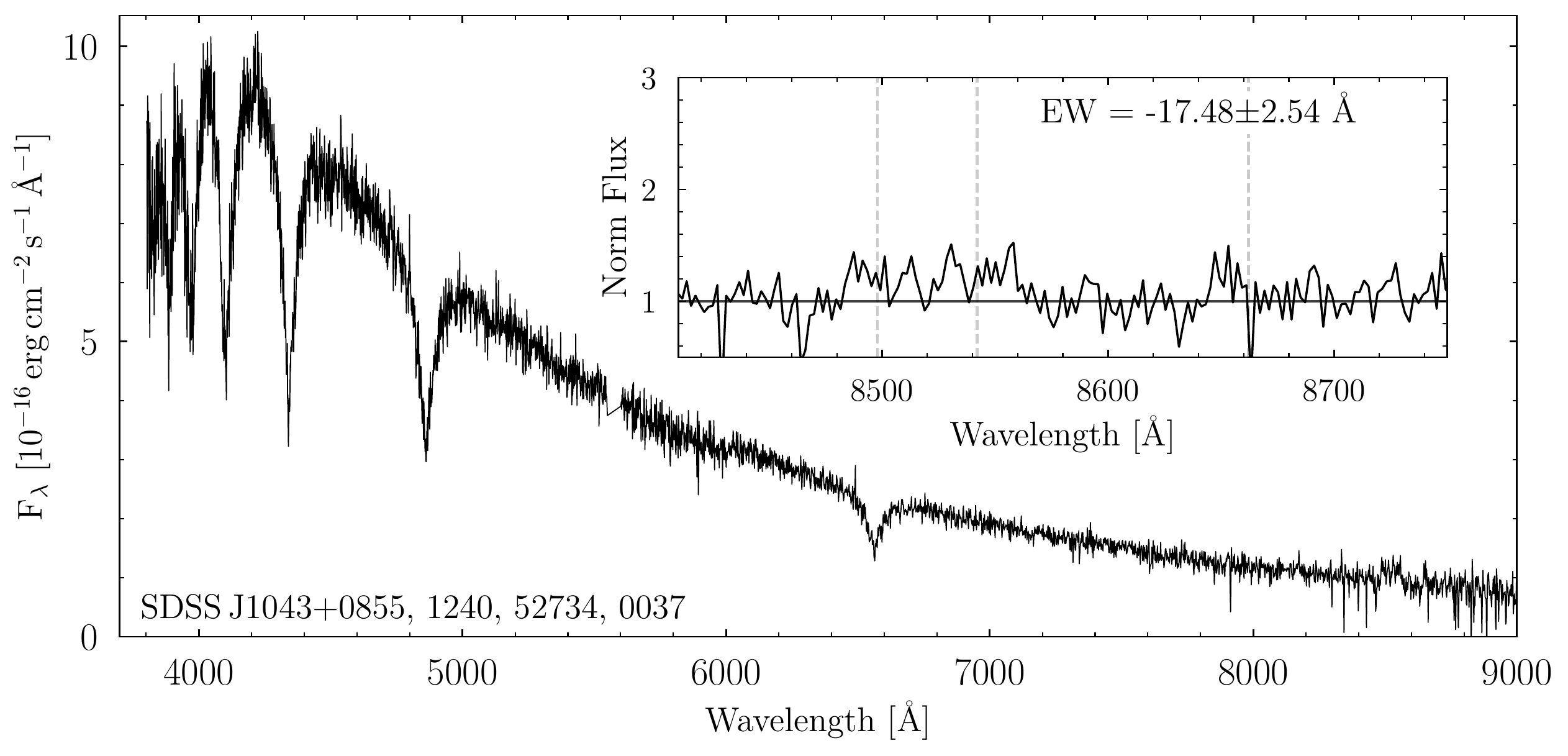}}
\caption{\label{f-normplot_appendix2} Continued.}
\end{figure} 

\renewcommand{\thefigure}{A\arabic{figure}}
\setcounter{figure}{0}

\begin{figure}
\centerline{\includegraphics[width=1\columnwidth]{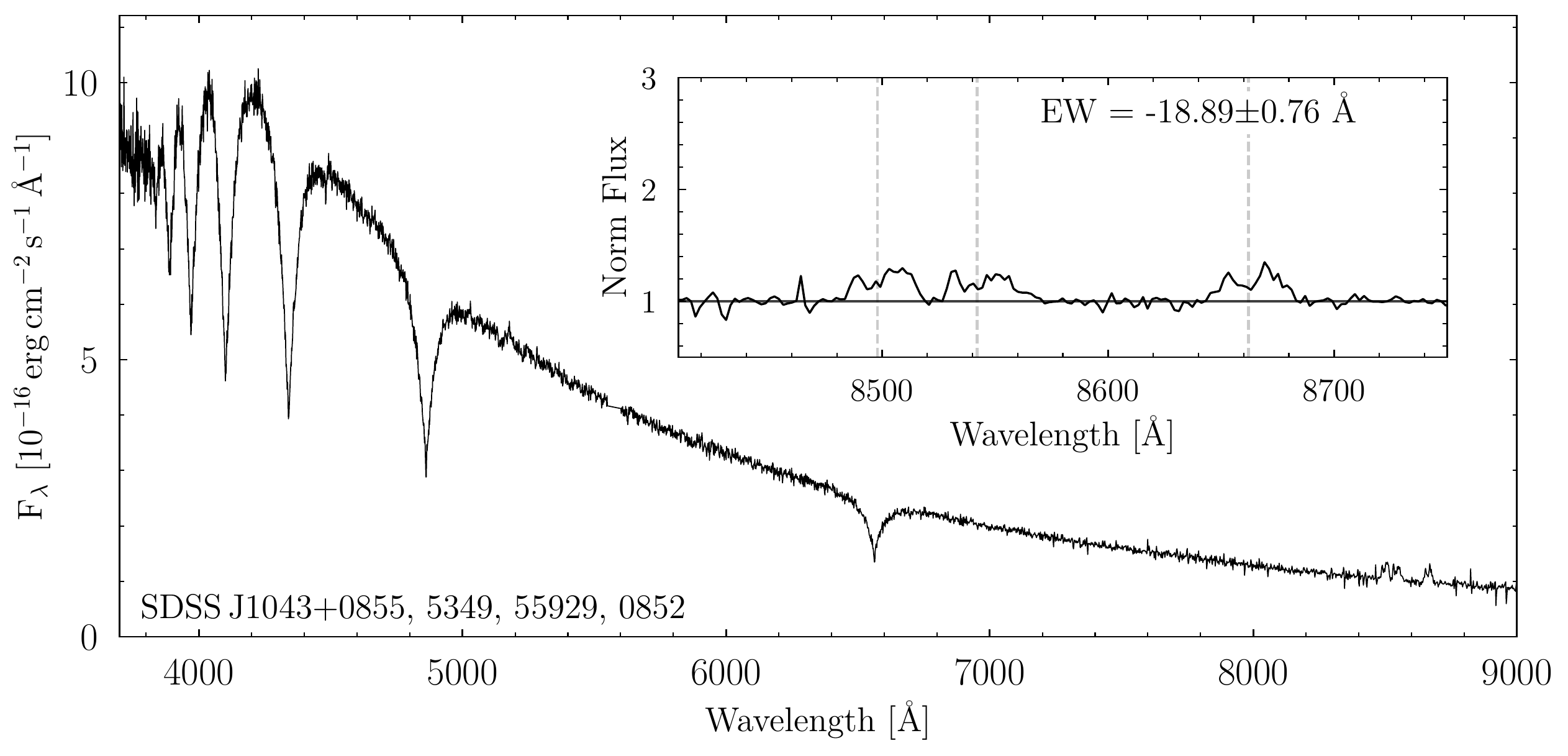}}
\centerline{\includegraphics[width=1\columnwidth]{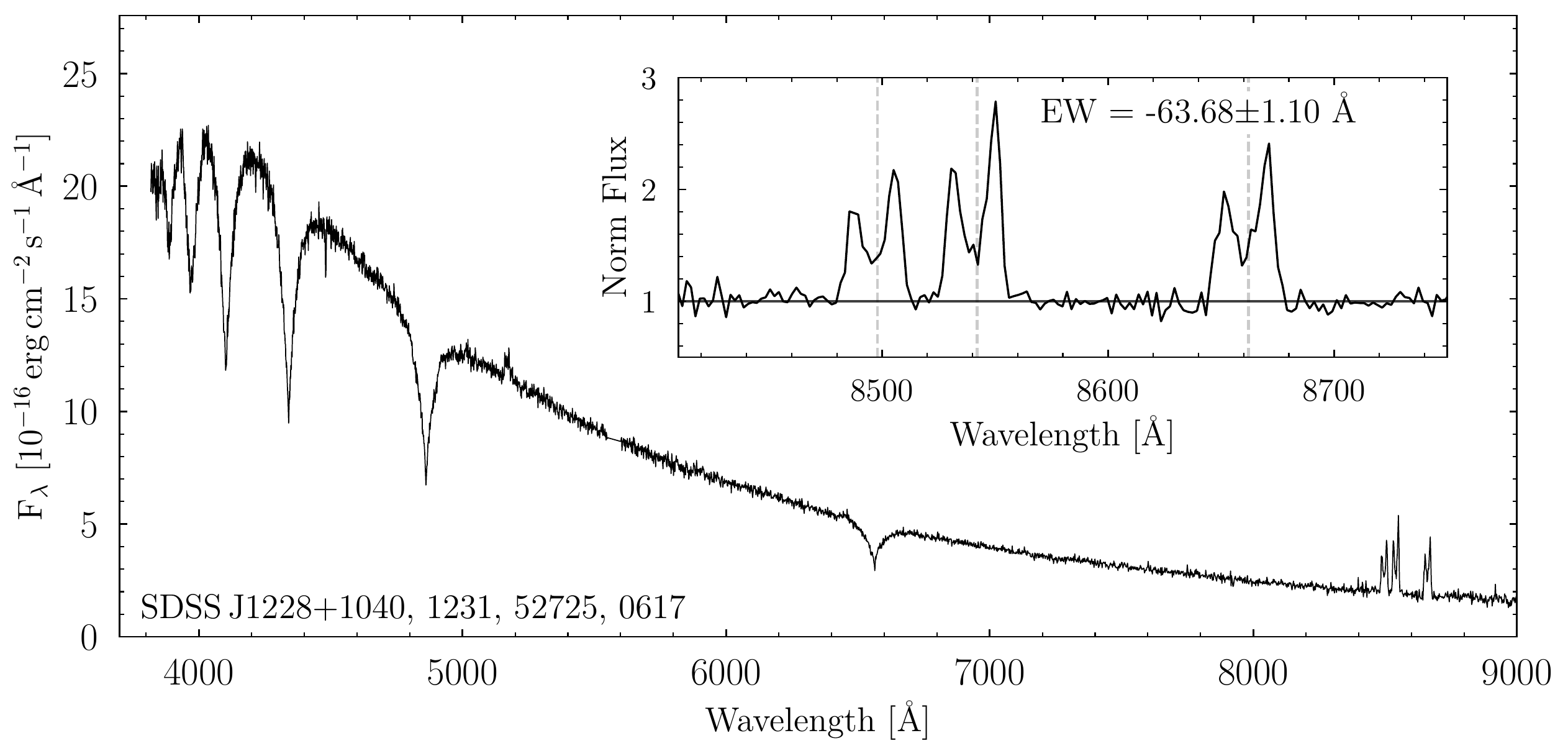}}
\caption{\label{f-normplot_appendix3} Continued.}
\end{figure} 

\renewcommand{\thefigure}{A\arabic{figure}}
\setcounter{figure}{0}

\begin{figure}
\centerline{\includegraphics[width=1\columnwidth]{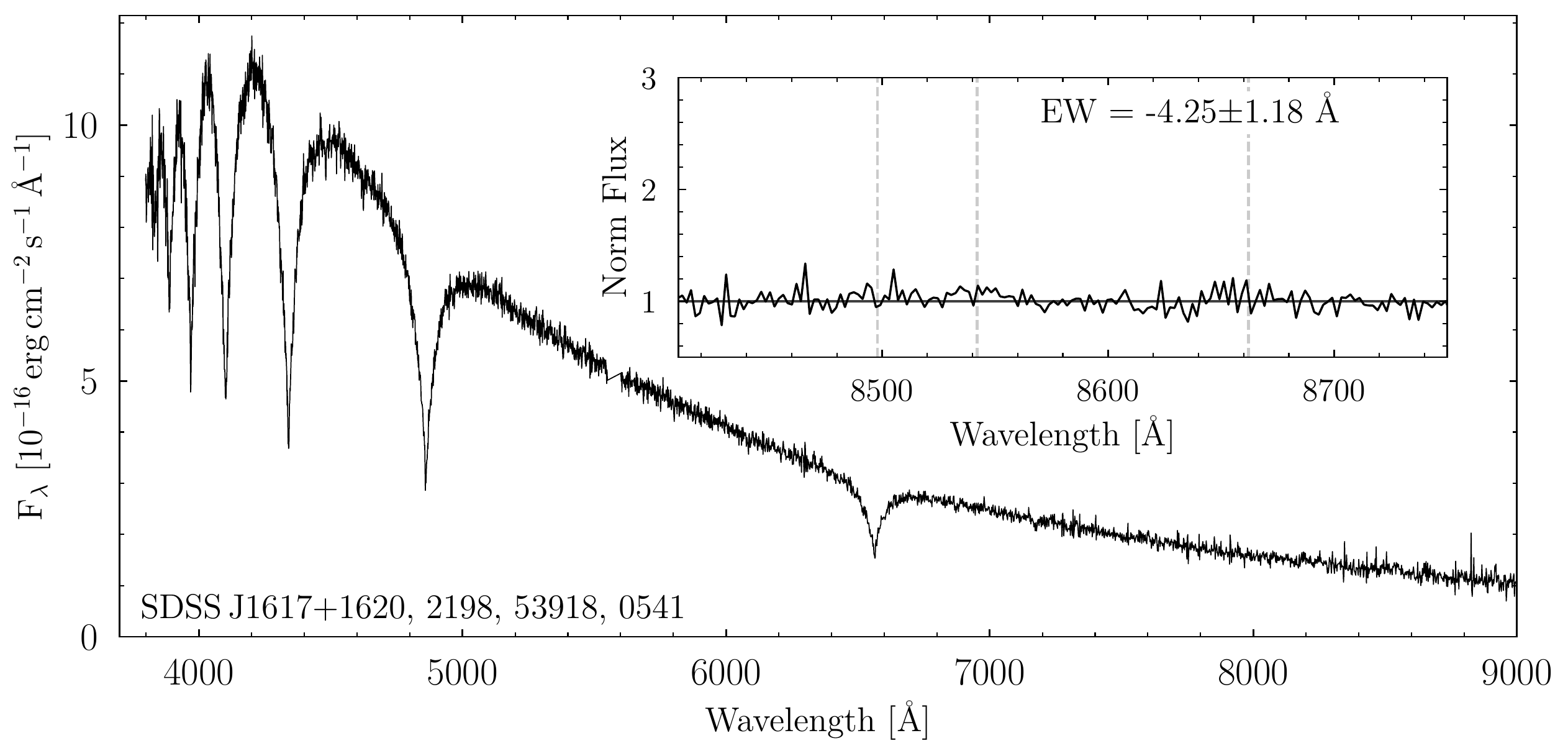}}
\centerline{\includegraphics[width=1\columnwidth]{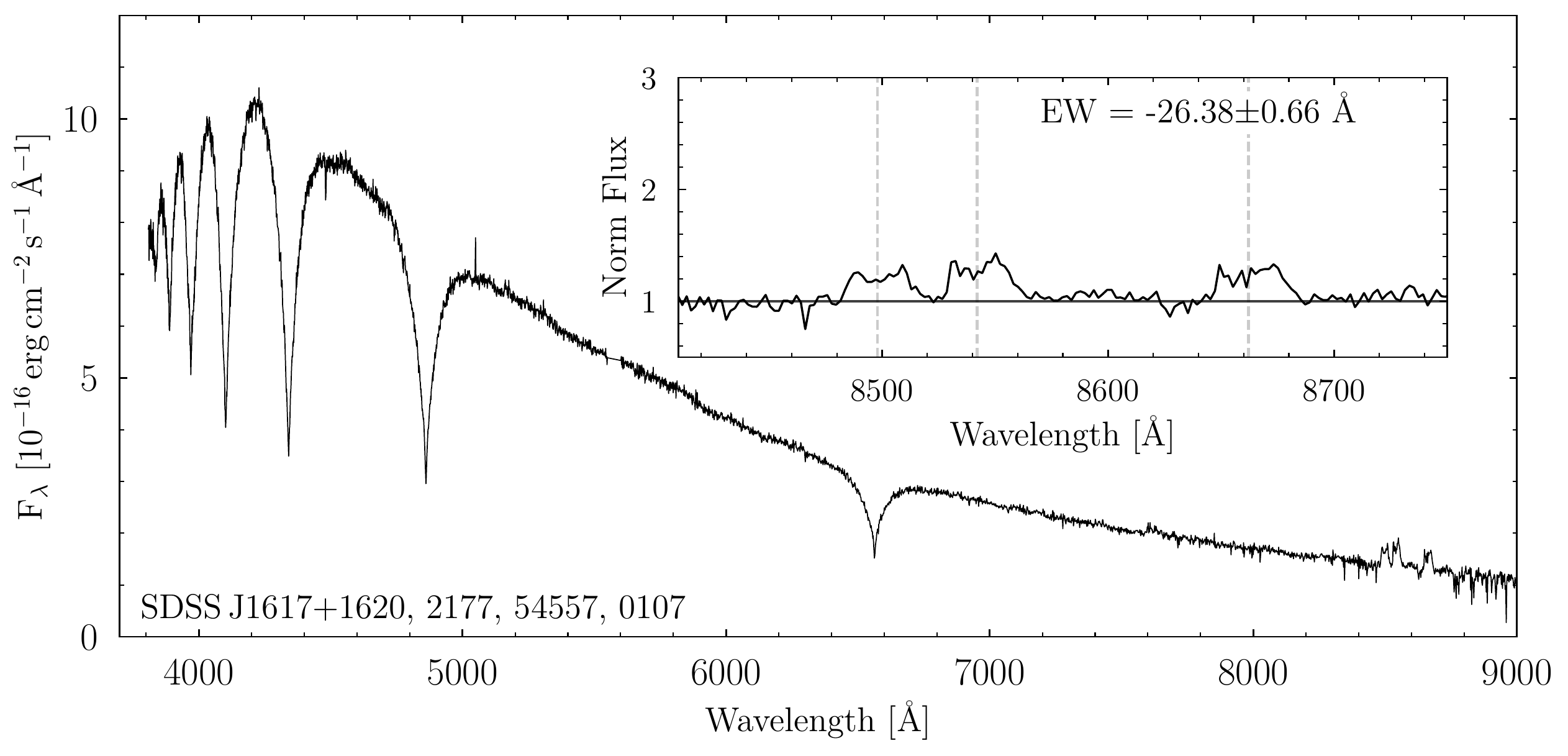}}
\caption{\label{f-normplot_appendix4} Continued.}
\end{figure}

\bsp

\label{lastpage}

\end{document}